\documentclass[10pt,journal,cspaper,compsoc]{IEEEtran}
\pdfoutput=1
\usepackage{graphicx}
\usepackage{color}
\ifCLASSOPTIONcompsoc
\else
\fi

\ifCLASSINFOpdf
\else
\fi

\begin{document}

\title{Input-Output Logic based Fault-Tolerant Design Technique for SRAM-based FPGAs}

\author{Aditya~Srinivas~Timmaraju*, Deshmukh~Aniket~Anand*, Mohammed~Amir~Khan, Zafar~Ali~Khan
\IEEEcompsocitemizethanks{\IEEEcompsocthanksitem  *Indicates equal contribution
\IEEEcompsocthanksitem Aditya Srinivas Timmaraju is with the Department of Electrical Engineering, Stanford University, CA - 94305\protect\\
E-mail address: adityast@stanford.edu
\IEEEcompsocthanksitem  Deshmukh Aniket Anand is with the Department of Electrical Engineering and Computer Science, University of Michigan, Ann Arbor, MI - 48109\protect\\
E-mail address: aniketde@umich.edu
\IEEEcompsocthanksitem  Dr. Zafar Ali Khan is with the Department of Electrical Engineering, Indian Institute of Technology, Hyderabad, AP, India - 502 205.\protect\\
E-mail address: zafar@iith.ac.in
\IEEEcompsocthanksitem  Mohammed Amir Khan is with the University Sains Malaysia, Pulau Penang, Malaysia.\protect\\E-mail address: amirpace@gmail.com} \\
\thanks{}}

%
%

\markboth{}
{Shell \MakeLowercase{\textit{et al.}}: Bare Demo of IEEEtran.cls for Computer Society Journals}
%

\IEEEcompsoctitleabstractindextext{%
\begin{abstract}
Effects of radiation on electronic circuits used in extra-terrestrial applications and radiation prone  environments need to be corrected. Since FPGAs offer flexibility, the effects of radiation on them need to be studied and robust methods of fault tolerance need to be devised. 
In this paper a new fault-tolerant design strategy has been presented. This strategy exploits the relation between changes in inputs and the expected change in output. Essentially, it predicts whether or not a change in the output is expected and thereby calculates the error. As a result this strategy reduces  hardware and time redundancy required by existing strategies like Duplication with Comparison (DWC) and Triple Modular Redundancy (TMR). The design arising from this strategy has been simulated and its robustness to fault-injection has been verified. Simulations for a 16 bit multiplier show that the new design strategy performs better than the state-of-the-art on critical factors such as hardware redundancy, time redundancy and power consumption.
\end{abstract}

}
\maketitle
\IEEEdisplaynotcompsoctitleabstractindextext

\IEEEpeerreviewmaketitle

\section{Introduction}
\IEEEPARstart{T}{he} ever-increasing desire of man to explore the extra-terrestrial space around him, to study and use it for mankind's better future, has given rise to a remarkable development on the space technology front in the last five decades. The technical think-tanks behind these developments have often had to face extraordinary challenges. One of these challenges is the vulnerability of electronic circuits used in space systems to powerful space radiations \cite{JoG}. Circuits used in space applications need to be protected from effects of such radiation. Closer home, radiation resistant circuitry is required for nuclear reactors and other radiation prone environments \cite{HoA}.

Full-custom hardware design (also known as Application Specific Integrated Circuits - ASICs) and Semi-custom hardware design (also known as Field Programmable Gate Arrays - FPGAs) are the two types of hardware that can be used for  radiation prone environments. Though ASICs offer the best performance for space applications, the complexity and cost involved are very high. Also, the functionality of ASICs is fixed and it cannot be altered. ASICs are very good against SEUs (Single Event Upsets) and they dominate the space technology market \cite{KCCH,ZhL}. On the other hand, FPGAs are reconfigurable devices. The most commonly used FPGAs are SRAM-based \cite{Cliv}. They offer the flexibility of changing the functionality, and most importantly, functionality can be modified on-field during a mission too \cite{StV,ZhL}. But, FPGAs are weak against SEUs \cite{AdA,ZMAP,AsT,Ol-et}. In the last decade, studies have been done to correct such errors, which occur due to radiations on SRAM-based FPGAs \cite{ZMAP,RCSK,YWZ,AdA,CBGTS}. These studies show that FPGAs have the potential to outperform ASICs in the space technology market. Thereby, it becomes necessary to study the effects of radiation on FPGAs and devise better fault-tolerant techniques.

The known correction techniques depend on replication (dual,triple) to provide protection \cite{Wake,Carm,KNHCR,KLCR} (and the references therein). In this paper we present a technique that provides protection based on logic between the inputs and output(s) of a gate-hence the name ``Input-Output Logic Based (IOLB)'' technique.

The remainder of the paper is organized is follows. Section \ref{sec2} reviews radiation effects and  existing techniques of fault-tolerance for SRAM based FPGAs. Section \ref{sec3} presents the proposed approach to fault-tolerant design. Section \ref{sec4} compares the proposed approach against the existing techniques. Section 5 concludes the paper.
\section{Review of Radiation Effects on SRAM-based FPGAs and Current Techniques}\label{sec2}
In this section we review the effects of radiation on SRAM-based FPGAs and the Current Techniques used to ameliorate these efffects.
\subsection{Radiation Effects on SRAM-based FPGAs}
Due to the effects of the earth's magnetic field and solar cycles, radiation doses change from location to location and time to time. So, a particular technology may be suitably adopted for a particular location and at a particular time \cite{BuK,SSJZ,Sau}. Space radiation effects are due to High Energy Electromagnetic Radiation and Particle radiations. This implies that Electromagnetic waves of high frequency viz., X-rays, Gamma rays and fast moving subatomic particles viz., neutrons, protons will affect SRAM-based FPGAs \cite{AdA,BDS,BuG}.

\begin{figure}[t]
  \caption{SEUs - An Illustration}\label{fig1}
       \includegraphics[height=70mm,width=70mm]{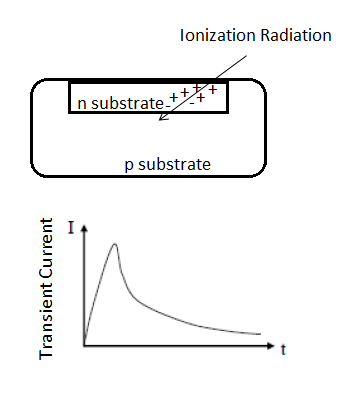}
\end{figure}

Broadly, the space radiation effects can be classified into Total Ionizing Dose (TID) and Single Event Upset (SEU). TID is the long-term damage done to the electronic circuitry due to electrons and protons. It is a permanent effect and it causes defects in the semiconductor lattice. Its effects are functional failures, leakage current and threshold shifts in CMOS \cite{MGBG}. SEU or soft error is a transient fault which may cause glitches i.e., current pulses, to propagate through the circuits. These glitches can be mitigated or corrected by design strategies.
      As shown in Fig. \ref{fig1}, in an SEU, ionization radiation loses its energy when it strikes the silicon in an electronic device due to the production of free electron-hole pairs.
Also, protons, neutrons and gamma rays can give rise to nuclear reactions when passing through a material. Ionization due to this generates large amounts of charge which is observed as a transient current pulse. Ohlsson et al. studied and analyzed the effect of neutrons in a Xilinx FPGA \cite{Ol-et}. In \cite{KNHCR}, Kastensmidt et al. point out that FPGAs are becoming more susceptible to neutrons as transistor size is decreasing and logic density increasing.

In SRAM-based FPGAs, customizable memory cells (SRAM cells), implement both the users combinational and sequential logic.

 When a SEU occurs in the combinational logic (synthesized in the  FPGA), it corresponds to a bit flip in one of the LUTs
cells or in the cells that control the routing. A SEU in an LUT memory cell modifies the implemented combinational
logic while an upset in the routing can connect or disconnect a wire in the matrix. The configuration bitstream's
next load corrects both these faults.

An SEU in the user sequential logic synthesized in the FPGA, has a transient effect
because the
next load corrects it. An SEU in the embedded block RAM has a permanent
effect and fault tolerance techniques must correct it.

A fault-tolerant system for a SRAM-based FPGAs, must cope with
the transient and permanent
effects of an SEU in the combinational logic, short
and open circuits in the design connections, and bit
flips in the flip-flops and memory cells\cite{KNHCR}.

Radiation tests on Xilinx FPGAs, for aerospace applications, have proven the need to use fault-tolerance schemes (for circuits)  \cite{FCB}. In \textcolor{black}{\cite{KNHCR}}, it is argued that protecting FPGAs by the use of redundancy is a lot more cost-effective as opposed to designing a new FPGA matrix of fault-tolerant elements. They move on to propose a fault-tolerant technique that employs time and hardware redundancy.

In \cite{Wake}, Wakerly attempted at applying Triple Modular Redundancy (TMR) concepts towards improving microcomputer reliability. In \cite{Carm}, Carmichael applied TMR methodology to Virtex FPGA series. For high-level SEU mitigation, the technique used most
often today to protect designs synthesized in the Virtex
architecture is based mainly on TMR combined with
scrubbing \cite{KNHCR}. In \cite{KLCR}, Kastensmidt et al. proposed two new schemes that reduced the redundancy from three in the TMR design to two. In this paper, we present the Input-Output Logic Based (IOLB) method that eliminates the need for even the existing dual redundancy.

\subsection{Triple Modular Redundancy}
\begin{figure}[t]
  \caption{TMR with a Majority Voter Block}\label{fig2}
       \includegraphics[height=36mm,width=70mm]{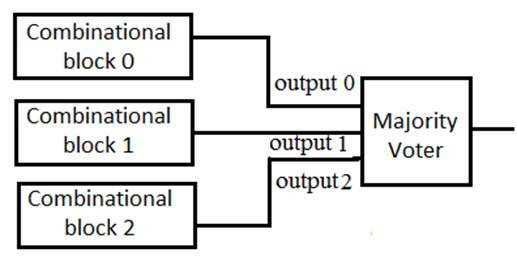}
\end{figure}

\begin{figure}[t]
  \caption{Majority Voter Circuit}\label{fig3}
       \includegraphics[height=37mm,width=70mm]{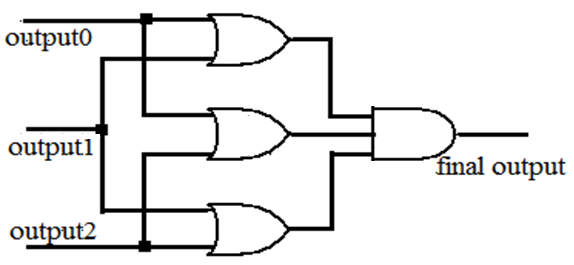}
\end{figure}

In TMR method, each pin, wire and block are triplicated and a majority voting is done to determine the correct output. The basic idea is depicted in Fig. \ref{fig2}. \textcolor{black}{An illustration of the majority voter circuit is presented in Fig. \ref{fig3}}. The area overhead in \textcolor{black}{TMR technique} is more than 3 times that of the standard circuit. It does not correct all the upsets. The upsets will accumulate if there is no extra logic for the refreshing. So typically, scrubbing is done (scrubbing is the process of reprogramming the FPGA periodically to ensure that faults do not accumulate) \cite{Wake,Carm}. Also, there are a variety of ways in which TMR can be applied to a circuit. In \cite{KLCR}, a comparison of the performance of these various ways is presented.

Note that scrubbing lets a system repair SEUs in the configuration
memory without disrupting operations (for correcting the voter logic). The
scrubbing cycle time depends on the configuration
clock frequency and the readback bitstream size \cite{KLCR}.

Overall, the TMR technique has
limitations such as high area overhead, three times
more input and output pins, and a significant increase
in power dissipation.
\subsection{Duplication with Comparison (DWC) and Time Redundancy}
\begin{figure}[t]
  \caption{DWC combined with Time Redundancy \cite{KLCR}}\label{fig4}
       \includegraphics[height=48mm,width=80mm]{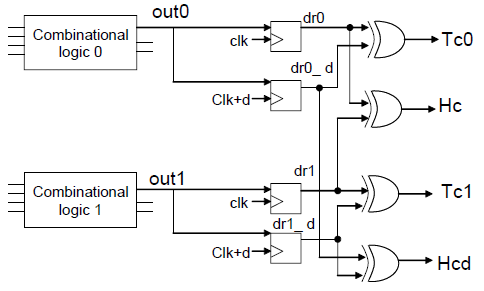}
\end{figure}

In this method, dual hardware and time redundancy concepts are used as presented in Fig. \ref{fig4} for arriving at the quartet (Tc0, Hc, Tc1, Hcd). The two redundant blocks used are labeled as combinational logic 0 (cl0) and combinational logic 1 (cl1). If an upset occurs in cl0, then Tc0 and Hcd will be '1' and Tc1 and Hc will be '0'. Similarly, an upset in cl1 can be detected when Tc1 and Hcd are '1' and Tc0 and Hc are '0'. Using this information, a state machine is designed to perform the voting of fault-free block. But when Tc0 is 0, Tc1 is 1, Hc is 0 and Hcd is 1, there is no way to predict the faulty block. For this reason, DWC with time redundancy may fail to correct stuck-at-zero and stuck-at-one faults as pointed in \cite{KLCR}. This led to the advent of DWC CED, which modifies time redundancy technique used in DWC with Time redundancy, to detect the permanent effect of an SEU.
\subsection{DWC with Concurrent Error Detection}
\begin{figure}[t]
  \caption{DWC-CED technique \cite{KLCR}}\label{fig5}
       \includegraphics[height=50mm,width=80mm]{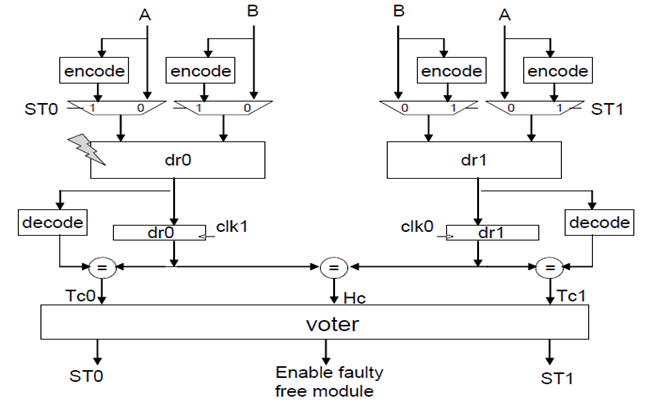}
\end{figure}

\begin{figure}[t]
  \caption{Voting the correct block \cite{KLCR}}\label{fig6}
       \includegraphics[height=60mm,width=80mm]{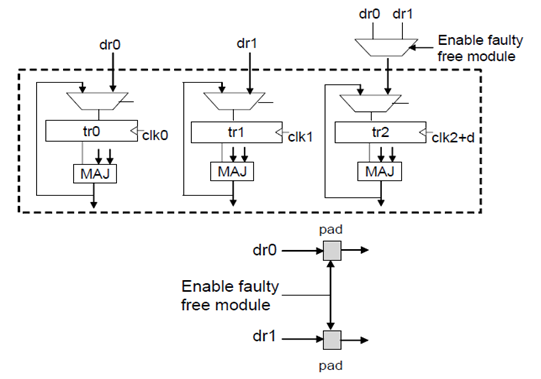}
\end{figure}

This method employs DWC along with Concurrent Error Detection (CED) to detect the location of the error.
         This method uses  encode and decode functions (as illustrated in Fig. \ref{fig5}) to re-compute the input operands. These functions are chosen in such a way that the output due to the re-computed operands differs from that of the original operands in the presence of an error. The voting circuit is presented in Fig. \ref{fig6}.
          Say there is an error in block 0. Then, Tc0 will be 1 and Hc will also be 1. From the state diagram of the voter circuitry \textcolor{black}{presented in \cite{KNHCR}}, it can be seen that it first enters the Upset Detection state as Hc is 1. From there, it enters the state dr1 is fault-free, since Tc0 is 1.
           \textcolor{black}{It is worth noting that there must not be upsets in more than one redundant module, including the detection and voting circuits for faithful functioning.}

           In both TMR and DWC-CED , scrubbing corrects
upsets in the user's combinational logic, and the CLB
flip-flops TMR scheme corrects upsets in the users
sequential logic. Scrubbing must be continuous to guarantee
that only one upset has occurred between two
reconfigurations in the design. As such, the scrubbing
rate should be fast enough to avoid the accumulation
of upsets in two different redundant blocks.
Upsets in the detection and voting circuit don't interfere
with the system's proper execution because the logic is
duplicated and the logics
latches are refreshed every
clock cycle \cite{KNHCR}.

\section{The Proposed Technique}\label{sec3}
The motivation of the proposed Input-Output Logic Based (IOLB) method is to reduce on the dual hardware redundancy and time redundancy of the state-of-the-art DWC-CED scheme. The new strategy exploits the relation between changes in inputs and expected changes in output. Such an approach to predicting an expected output is novel and has not been reported hitherto. Our strategy does not require explicit duplication or triplication of the combinational logic block to detect and correct an error.

The proposed design methodology assesses if the output is expected to change following a change in the input(s). These signals (changes in inputs and change in output) are used along with appropriately designed logic to generate the error signal, which can then be XOR-ed with the output signal to yield the error-free output. Like in TMR and DWC-CED, scrubbing needs to be employed to correct upsets in combinational logic and TMR needs to be employed to correct upsets in sequential logic.
\subsection{NOT Gate}
We shall now consider the case of a NOT gate to illustrate the mechanism involved in designing the IOLB correction circuit. Say $A$ is its input and $B$ is its output. In an error-free scenario, if $A$ is '1', then $B$ would be '0'. However, say $A$ is now changed to '0' (due to a SEU)  and there is no accompanied change in $B$ (that is, it stays at '0'). Then, that will mean that there is an error, since, in a NOT gate, a change in input is expected to bear a change in the output too.

Let us now examine a couple of other cases of input-output pair transitions. If the pair (format: 'AB') changes from '01' to '00', this means that the output change has occurred without there being a change in the input. This is again unexpected behavior for a normal NOT gate. Similar would have been the case, if the pair transitioned from '10' to '11'.

These above relations form the central idea to designing the error correcting circuit. Suppose $A_{c}$ is change in the input, $B_{c}$ is the change in the output and $E$ is the error signal. We can arrive at the following truth table from the above inferences drawn about relations between the change in input and the change in output.
\begin{table}[t]
\caption{NOT Gate: Generating E} 
\centering 
\begin{tabular}{|c| c| c|} 
\hline
A$_{c}$ & B$_{c}$ & E  \\ [1ex] 
\hline 
0 & 0 & 0 \\ 
0 & 1 & 1 \\
1 & 0 & 1 \\
1 & 1 & 0 \\[1ex] 
\hline 
\end{tabular}
\label{tab1} 
\end{table}
\begin{figure}[t]
  \caption{Block diagram of IOLB NOT gate }\label{fig7}
  \centering
       \includegraphics[height= 38mm,width=65mm]{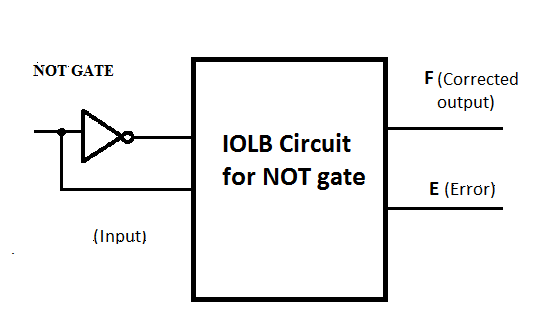}
\end{figure}
\begin{figure}[t]
  \caption{Circuit diagram of IOLB NOT gate }\label{fig8}
  \centering
       \includegraphics[height= 38mm,width=80mm]{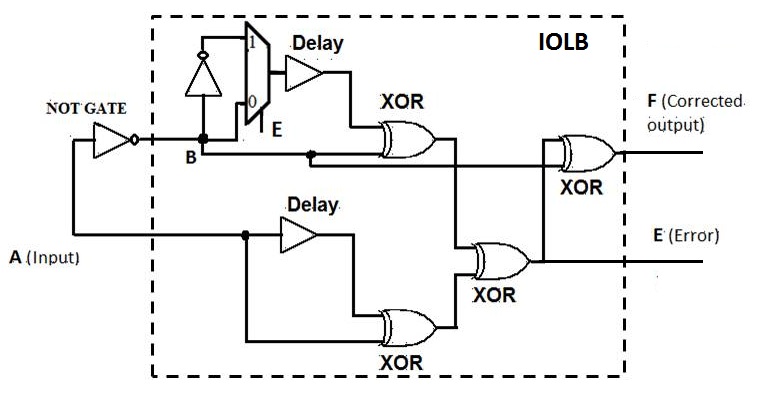}
\end{figure}
       In Table \ref{tab1}, in the first case, if there is no change in the input and the output (syndrome '00'), there is no error. If there is no change in the input but there is a change in the output (syndrome '01'), there is an error. Similarly, rest of the syndromes can be tracked.

       For the computation of changes in variables, we take the XOR of a variable with a delayed version of itself, thereby giving us '1' if there has been a change. The value of the delay is arbitrary. While trying to compute $B_{c}$ from $B$, we need to take care of the possibility of an error having occurred in the NOT gate. If we just perform XOR of delayed and current values of B, we will get an erroneous $B_{c}$ if the NOT gate is error-affected (since $B$ would be erroneous). Hence, if the error signal $E$ is '1' (indicating that the NOT gate is affected), we take XOR with the NOT of delayed B for the computation of $B_{c}$. A block diagram of  IOLB NOT gate is illustrated in Fig. \ref{fig7} and the IOLB circuit for NOT gate is shown in Fig \ref{fig8}.
\subsection{Exclusive OR (XOR) Gate}
We now consider the case of a two-input XOR gate. Say $A$ and $B$ are its inputs and $S$ is its output. In an error free scenario, if both $A$ and $B$ are same, then $S$ would be '0' and if both are different then $S$ would be '1'. However, in case both $A$ and $B$ are same and $S$ is '1' or in case both $A$ and $B$ are different and $S$ is '0', that would mean there is an error. Such a case will arise when an expected output change does not succeed a change in an input/ when an unexpected output change occurs, even without a change in the inputs. We shall now look at specific instances of the above cases.

Say the current state (format: 'ABS') is '000'. Suppose a transition occurs from '000' to '100', this is unexpected because, in an XOR gate, output is expected to change following a change in exactly one of the input. Suppose a transition occurs from '000' to '001', this is again unexpected, since in an XOR gate, the output cannot change without a change in its inputs.

Henceforth, we shall use c-subscripted symbols to denote changes in variables. For instance, X$_c$ denotes change in the variable X. In other words, X$_c$='1' means there has been a change in X and X$_c$='0' means there hasn't been a change in X.

In table 2, in the first case, there is no change in either of the inputs and there is no change in the output. Hence, the error (denoted as $E$) is zero. In the second case, there is no change in either of the inputs, but there is a change in the output. As has been discussed above, this is unexpected and hence there is an error ($E$ = '1'). In the third case, exactly one of the inputs has changed and hence the output is expected to change. But it hasn't changed (since $S_c$ = '0') and hence there is an error ( $E$ = '1'). Rest of the cases can be seen tracked from truth table in a similar fashion.

\begin{table}[t]
\caption{XOR Gate: Generating E} 
\centering 
\begin{tabular}{|c| c| c| c|} 
\hline
A$_{c}$ & B$_{c}$ & S$_{c}$ & E  \\ [0.5ex] 
\hline 
0 & 0 & 0 & 0 \\ 
0 & 0 & 1 & 1\\
0 & 1 & 0 & 1 \\
0 & 1 & 1 & 0 \\
1 & 0 & 0 & 1\\
1 & 0 & 1 & 0 \\
1 & 1 & 0 & 0\\
1 & 1 & 1 & 1\\[1ex] 
\hline 
\end{tabular}
\label{table:nonlin} 
\end{table}
\begin{figure}[t]
  \caption{Block Diagram of IOLB XOR gate }
  \centering
       \includegraphics[height= 38mm,width=65mm]{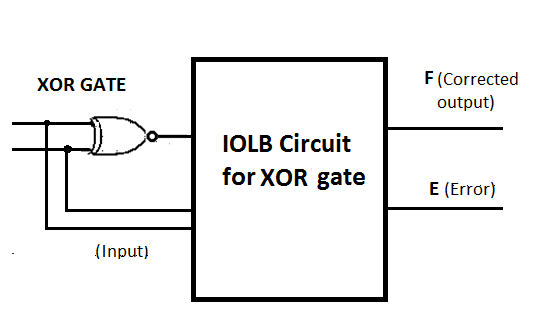}
\end{figure}

\begin{figure}[t]
  \caption{Circuit Diagram of IOLB XOR gate }
       \includegraphics[height= 50mm,width=80mm]{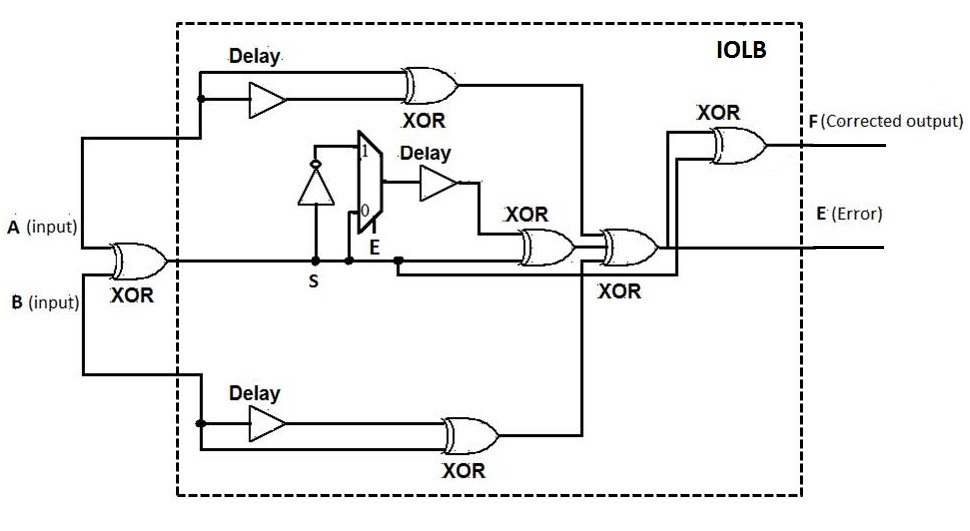}
\end{figure}
From Table 2, it can be inferred that $E$ = $A_c$ (xor) $B_c$ (xor) $S_c$.

For arriving at changes in variables, we take the XOR of a variable and its delayed version. However, like in the case of NOT gate, the computation of $S_c$ is not straightforward. We resolve the problem in the same way as was done in the case of NOT gate: in the computation of $S_c$, we use NOT of delayed $S$ if $E$ = '1'.

For producing the fault-free output, the error signal E is XOR-ed with S. A complete picture is illustrated in Fig 9 and the IOLB circuit for XOR gate is shown in Fig 10.

\begin{figure}[t]
  \caption{Error correction: TMR for a XOR gate}\label{fig_tmr_xor}
  \centering
       \includegraphics[height=40mm,width=63mm]{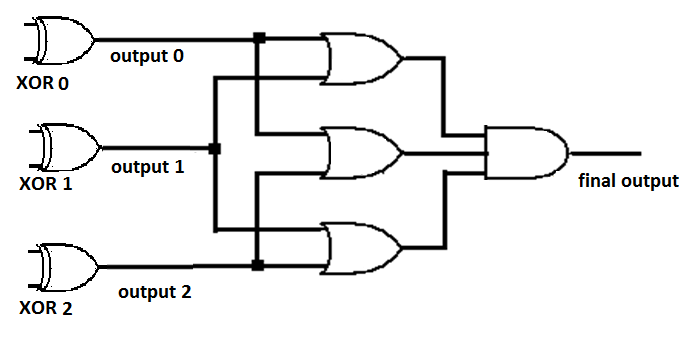}
\end{figure}

\begin{figure}[t]
  \caption{DWC-CED for a XOR gate [20]}\label{fig_dwc_xor}
       \includegraphics[height=60mm,width=90mm]{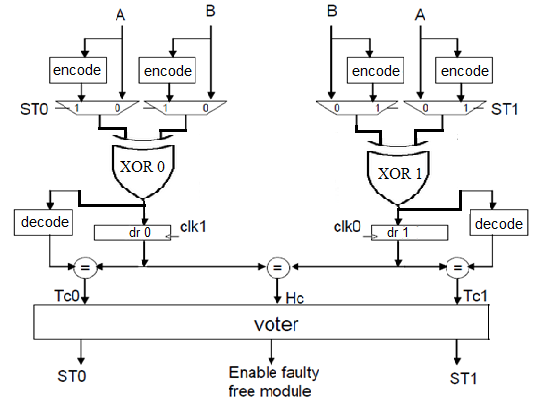}
\end{figure}

\textcolor{black} {One may think that in IOLB, for a single XOR gate, there is an addition of 5 XOR gates and 1 multiplexer. But it should be noted that TMR would have a six-fold increase (refer Fig. \ref{fig_tmr_xor}) in resources when looked at for a single XOR gate. TMR also requires triplication of inputs. For the DWC-CED method too, the resources required would be 5 XOR Gates + 4 Encode, 2 Decode Blocks + 4 multiplexers, 2 Flip-Flops and additional voter circuitry (refer Fig. \ref{fig_dwc_xor}).}

\subsection{General Procedure}
{Consider any logic gate with $X_1, X_2, X_3...X_n$ as inputs and $Y$ as the output. Like earlier, let $A_c$ denote change in a variable $A$. If $A_c$ is 1, there has been a change in $A$ and if $A_c$ is 0, there has been no change in $A$. We first obtain the changes in the input variables $(X_{1,c}, X_{2,c}, X_{3,c}...X_{n,c})$ (labeled as change variables) by XOR-ing $X_1, X_2, X_3...X_n$ with their delayed versions. Likewise, we also obtain $Y_c$ from $Y$. Then, we analyze all possible cases of input-output transitions. There would be $2^{2n+2}$ cases, because given an input state, there are $2^n$ possible transitions (since there are 'n' change variables) and the number of input states is itself $2^n$ (since there are 'n' input variables). Also, the output $Y$ and its change variable $Y_c$ contribute to 2$^2$ states.}

{\textcolor{black}{The next step involves generating the error variable $E$, which indicates whether or not there is an error. We arrive at a truth table using $X_1, X_2, X_3...X_n, X_{1,c}, X_{2,c}, X_{3,c}...X_{n,c}, Y, Y_c$ (labeled as error inputs) and estimate the error variable $E$ in each of the $2^{2n+2}$ cases. After the error variable $E$'s column in the truth table is completed, $E$ is expressed as a function of the error inputs. Sometimes, as witnessed in the above cases of NOT and XOR, the state where the transition began does not matter in generating $E$, only the transitions itself matter. So, we look for such potential redundancies and formulate an expression for the error variable $E$. The general expression can be presented as,\\}

$E = f(X_1, X_2, X_3...X_n, X_{1,c}, X_{2,c}, X_{3,c}...X_{n,c}, Y, Y_c)$\\

{\textcolor{black}{After the error variable $E$ is obtained, the output $Y$ is XOR-ed with $E$ to generate the error-free output $F$.}

\section{Comparison}\label{sec4}
A 16-bit multiplier has been implemented using the IOLB strategy presented above. The IOLB circuits for AND and OR gates also have been arrived at, using the above strategy. Each of the logic gates in the cascaded multiplier were replaced with their full IOLB counterparts. We evaluated our method by testing it against fault-injection and the percentage of corrected faults was found to be 100\%. Just like in \cite{KNHCR}, we utilized 4x1 fault-injection multiplexers for emulating stuck-at-zero and stuck-at-one faults. A comparison of the usage of resources by different methods is presented in Table 3.

As can be seen from Table 3, the proposed design strategy requires lower hardware resources than do DWC-CED and TMR. Delay and power consumption factor could not be compared as we could not gather the clock frequency that was used in the study presented in \cite{KNHCR}. However, we argue that our technique will perform better on delay, because there is no explicit time redundancy like in DWC-CED. Also, IOLB will do better at power consumption because of lesser requirement of both hardware resources and time.

\begin{table}[t]
\caption{Comparison of utilized resources} 
\centering 
\begin{tabular}{|c| c| c| c|} 
\hline
Technique & I/O Pads & 4-input LUTs & Flip-flops  \\ [0.5ex] 
\hline 
None & 67 & 495 & 32\\
\hline
TMR & 201 & 1709 & 96 \\
\hline
DWC-CED \cite{KNHCR} & 169 & 1706 & 162 \\
\hline
IOLB & \textcolor{black}{137} & 1536 & 96\\[1ex] 
\hline 
\end{tabular}
\label{table:nonlin} 
\end{table}

\begin{table}[t]
\caption{TMR Technique} 
\centering 
\begin{tabular}{|c| c| c| c|} 
\hline
M$_{1}$ & M$_{2}$ & M$_{3}$ & Output  \\ [0.5ex] 
\hline 
0 & 0 & 0 & Faithful \\ 
0 & 0 & 1 & Faithful\\
0 & 1 & 0 & Faithful \\
0 & 1 & 1 & Not faithful \\
1 & 0 & 0 & Faithful\\
1 & 0 & 1 & Not faithful \\
1 & 1 & 0 & Not faithful\\
1 & 1 & 1 & Not faithful\\[1ex] 
\hline 
\end{tabular}
\label{table:nonlin} 
\end{table}

\begin{table}[!t]
\caption{DWC-CED Technique} 
\centering 
\begin{tabular}{|c| c| c|} 
\hline
M$_{1}$ & M$_{2}$ & Output  \\ [0.5ex] 
\hline 
0 & 0 & Faithful \\ 
0 & 1 & Faithful\\
1 & 0 & Faithful \\
1 & 1 & Not faithful \\ [1ex] 
\hline 
\end{tabular}
\label{table:nonlin} 
\end{table}

\begin{table}[!t]
\caption{IOLB Technique} 
\centering 
\begin{tabular}{|c| c|} 
\hline
M$_{1}$ & Output  \\ [0.5ex] 
\hline 
0 & Faithful \\ 
1 & Faithful \\ [1ex] 
\hline 
\end{tabular}
\label{table:nonlin} 
\end{table}

\textcolor{black} {In the event of an SEU, all the three schemes are error free. However, in TMR there is a possibility that error can occur in two or more replications. Similarly, in DWC-CED, there is a possibility that error occurs in both the replications while in our scheme error can occur at most in one module. While the scheme of this paper will correct possible errors, DWC and TMR will fail for two or more errors. This is elaborated in the following analysis.}

In the Tables 4, 5 and 6, M$_{i}$ indicates the i$\mathrm{^{th}}$ module. If M$_{i}$ is 0, then it is fault-free and if M$_{i}$ is 1, there is a fault in it. In TMR, a module is triplicated and a faithful functioning is expected if two of the three instances are fully free from error. In Table 4, have identified eight possible scenarios for TMR, each of which could occur with equal probability.

As can be seen from Table 4, TMR works in 4 out of the 8 possible cases. This is because two of the three instances of the module should be fault-free for a faithful functioning. From Table 5, it can be inferred that DWC-CED works in 3 out of the 4 possible cases. This is because DWC-CED works in all cases except when both the instances of the module (i.e., both M$\mathrm{_{1}}$ and M$\mathrm{_{2}}$) are affected.

\begin{table}[!t]
\caption{Theoretical Analysis: A Summary} 
\centering 
\begin{tabular}{|c| c|} 
\hline
Technique & Probability of faithful functioning  \\ [0.5ex] 
\hline 
TMR & 1/2 \\ 
DWC-CED & 3/4\\
IOLB & 1 \\ [1ex] 
\hline 
\end{tabular}
\label{table:nonlin} 
\end{table}

We present a similar analysis in Table 6 for IOLB technique. The IOLB technique uses only one instance of the original module. If there is a fault in the module, then the IOLB circuit corrects it. So, in both cases, a faithful output is obtained. We thus conclude from our above analysis that IOLB is the least likely to be affected when compared with TMR and DWC-CED. A summary of the findings from the above analysis is presented in Table 7.

\section{Conclusion \& Future Directions}\label{sec5}

In this paper, we have presented a new strategy for fault-tolerant design that performs better than the state-of-the-art. The proposed strategy essentially provides a mechanism to predict whether or not a change in the output is expected, as a function of changes in inputs. For this reason, the utility of this circuit is not only confined to fault-tolerant design but also relevant to applications that require information about whether or not changes in inputs would result in changes in output(s). Directions for future work include looking at other application domains where the proposed strategy can be applied to good effect.

\ifCLASSOPTIONcaptionsoff
  \newpage
\fi



\begin{thebibliography}{1}


\bibitem{JoG}
A.H. Johnston and S.M. Guertin, ``The effects of space radiation on linear integrated circuits,'' \textit{Proc. Aerospace Conf.,} vol. 5, pp. 363--369, 2000.

\bibitem{Cliv}
Clive Maxfield, ``FPGA Architectures,'' \textit{FPGAs: Instant Access}, Oxford, UK: Newnes Publication, July 2008, ch. 2 , pp. 14.


\bibitem{StV}
L. Sterpone and M. Violante, ``Analysis of the robustness of the TMR architecture in SRAM-based FPGAs,'' \textit{IEEE Trans. Nuclear Science,} vol. 52, no. 5, pp. 1545--1549, 2005.

\bibitem{KCCH}
R. Koga, W. R. Crain, K. B. Crawford, S. J. Hansel, S. D. Pinkerton, and T. K. Tsubota, ``The Impact of ASIC Devices on the SEU Vulnerability of Space-Borne Computers,''  \textit{IEEE Trans. Nuclear Science}, vol. 39, no. 6, pp. 1685--1692, Dec. 1992.

\bibitem{AdA}
Philippe Adell and Greg Allen, ``Assessing and Mitigating Radiation Effects in Xilinx FPGAs,'' Jet Propulsion Laboratory, \textit{California Institute of Technology, JPL Publication,} 2008, 08-9 2/08..


\bibitem{BDS}
J. Barth, C. Dyer, and E. Stassinopoulos, ``Space, Atmospheric, and Terrestrial Radiation Environments,'' \textit{IEEE Trans. Nuclear Science,} vol. 50, no. 3, pp. 466--482, June 2003.

\bibitem{ZMAP}
Hamid R. Zarandi, Seyed Ghassem Miremadi, Costas Argyrides, Dhiraj K. Pradhan, ``\textcolor{black}{Fast} SEU Detection and Correction in LUT Configuration Bits of SRAM-based FPGAs,''  \textit{Proc. IEEE Int'l Parallel and Distributed Processing Symp. (IPDPS '07),} pp. 1--6, 2007.

\bibitem{RCSK}
E. Syam Sundar Reddy, Vikram Chandrasekhar, \textcolor{black}{M. }Sashikanth and V. Kamakoti,``Detecting SEU-caused Routing Errors in SRAM-based FPGAs,''  \textit{Proc. 18th Int'l Conf. VLSI Design (VLSID '05),} pp. 736--741, 2005.

\bibitem{YWZ}
Wenlong Yang, Lingli Wang, Xuegong Zhou, ``CRC Circuit Design for SRAM-Based FPGA Configuration Bit Correction,'' \textit{Proc. 10th IEEE Int'l Conf. Solid-State and Integrated Circuit Technology (ICSICT '10),} pp. 1660--1664, 2010.

\bibitem{AsT}
Ghazanfar-Hossein Asadi and Mehdi Baradaran Tahoori, ``Soft Error Mitigation for SRAM-Based FPGAs,'' \textit{Proc. 23rd IEEE VLSI Test Symp.,} pp. 207--212, 2005.

\bibitem{Ol-et}
M. Ohlsson , P. Dyreklev, K. Johansson, and P. Alfke, ``Neutron single event upsets in SRAM-based FPGAs,'' \textit{Proc. IEEE Nuclear Space Radiation Effects Conf. (NSREC '98),} pp. 1--4, 1998.


\bibitem{CBGTS}
L.V. Cargnini, R. Brum, Y. Guillemenet, L. Torres, and G. Sassatelli, ``Improving the Reliability of a FPGA using Fault-Tolerance Mechanism Based on Magnetic Memory (MRAM),'' \textit{Proc. Int'l Conf. Reconfigurable Computing and FPGAs (ReConFig '10),} pp. 150--155, 2010.

\bibitem{BuG}
N. J. Buchanan and D. M. Gingrich, ``Proton Radiation Effects in XC4036XLA Field Programmable Gate Arrays,'' \textit{IEEE Trans. Nuclear Science,} vol. 50, no. 2, pp. 263--271, 2003.

\bibitem{BuK}
B. Ravinarayana Bhat, Nagesh Upadhyaya,
 and Ravi Kulkarni, ``Total Radiation Dose at Geostationary Orbit,'' \textit{IEEE Trans. Nuclear Science,} vol. 52, no. 2, pp. 530--534, 2005.

\bibitem{SSJZ}
T. Shelfer, E. Semones, S. Johnson, N. Zapp, M. Weyland, F. Riman, J. Flanders, M. Golightly and G. Smith, ``Active Radiation Monitoring on the International Space Station,''
 \textit{Presented at the 32nd Int'l Conf. Environment Systems (ICES '02),} 2002.

\bibitem{Sau}
H.H. Sauer, ``Notes on the Natural Radiation Hazard at Aircraft Altitudes,'' 1 Oct. 2007; www.swpc.noaa.gov/info/RadHaz.htm

\bibitem{MGBG}
D. M. MacQueen, D. M. Gingrich, N. J. Buchanan and P. W. Green, ``Total Ionizing Effects in a SRAM-Based FPGA,''  \textit{Proc. Radiation Effects Data Workshop,} pp. 24--29, July 1999.

\bibitem{Wake}
J. F. Wakerly, ``Microcomputer Reliability  Improvement  Using Triple-Modular Redundancy,'' \textit{Proc. IEEE,} vol. 64, no. 6, pp. 889--895, 1976.

\bibitem{Carm}
C. Carmichael, ``Triple Module Redundancy Design Techniques for Virtex Series FPGA,'' Xilinx Application Notes 197, v1.0, Mar. 2001, pp. 137.

\bibitem{KLCR}
F.L. Kastensmidt, L. Carro, and R. Reis, \textit{Fault-tolerance Techniques for SRAM-based FPGAs,} Springer, 2006.


\bibitem{ZhL}
Zhang and Liu, ``Synthetic Analysis on Space Radiation Tolerance Techniques in ASICs and FPGAs,'' \textit{Proc. Int'l Conf. System Science, Engineering Design and Manufacturing Informatization (ICSEM '11),} vol. 2, pp. 305--310, Oct. 2011.

\bibitem{KNHCR}
F.L. Kastensmidt, G. Neuberger, R.F. Hentschke, L. Carro, and R. Reis, ``Designing Fault-Tolerant Techniques for SRAM-Based FPGAs,'' \textit{IEEE Design \& Test of Computers,} vol. 21, no. 6, pp. 552--562, 2004.


\bibitem{FCB}
E. Fuller, M. Caffrey, P. Blain, C. Carmichael, N. Khalsa, and A. Salazar, "Radiation test results of the virtex FPGA and ZBT SRAM for space based reconfigurable computing," \textit{Proc. Military and Aerospace PLD Conf.,} Sep. 1999, pp. 1--8.

\bibitem{KLSC}
F.L. Kastensmidt, L. Sterpone, L. Carro, and M.S. Reorda, "On the optimal design of triple modular redundancy logic for SRAM-based FPGAs," \textit{Proc. Conf. Design, Automation and Test in Europe (DATE '05),} vol. 2, pp. 1290--1295, 2005.

\bibitem{HoA}
 A. G. Holmes-Siedle and L. Adams, \textit{ Handbook of Radiation Effects }, Oxford University Press, England 2002.




\end{thebibliography}
\end{document}